\documentclass{aa}  

\usepackage{graphicx}
\usepackage{txfonts}
\usepackage{lipsum}
\usepackage{subcaption}         
\usepackage{lscape}             
\usepackage{placeins}           
                                
\usepackage[normalem]{ulem}
\usepackage{xcolor}
\newcommand{\martiniii}[1]{\textcolor{blue}{}}

\newcommand{\tcore}{$\tilde{T}_{c}$}
\newcommand{\tb}{$\tilde{T}_{b}$}
\newcommand{\nscool}{\texttt{NSCool}}

\newcommand{\mxb}{MXB~1659$-$29}
\newcommand{\mxbbrief}{MXB 1659}
\newcommand{\mxbi}{MXB 1659 (OI)}

\newcommand{\ks}{KS 1731$-$260}
\newcommand{\ksbrief}{KS 1731}
\newcommand{\xte}{XTE J1701$-$462}
\newcommand{\xtebrief}{XTE J1701}
\newcommand{\igr}{IGR J17480$-$2446 (Ter5 X-2)}
\newcommand{\igrS}{IGR J17480$-$2446}
\newcommand{\igrbrief}{Ter5 X-2}

\newcommand{\superfluids}{$^{1}\mathrm{S}_{0}$}
\newcommand{\superfluidp}{$^{3}\mathrm{P}_{2}$}

\newcommand{\dingi}{CDBonn SRC}
\newcommand{\dingii}{CDBonn SRC+P}
\newcommand{\dingiii}{Av18 SRC}
\newcommand{\dingiv}{Av18 SRC+P}
\newcommand{\dingv}{N3LO SRC}
\newcommand{\dingvi}{N3LO SRC+P}

\begin{document}

   \title{On the neutron star cooling in transient X-ray sources as probes of the neutron $^{3}\mathrm{P}_{2}$ pairing}


   \author{M. Nava-Callejas\inst{1}\fnmsep\thanks{Corresponding author: martin.javier.nava.callejas@ulb.be}
        \and N. Chamel\inst{1}
        }

   \institute{Institut d'Astronomie et d'Astrophysique, Université Libre de Bruxelles, 1050 Bruxelles, Belgique}

   \date{Received September 30, 20XX}
 
  \abstract
   {The observed rapid cooling of the isolated neutron star in Cassiopeia A is generally explained by the existence of neutron \superfluidp{} superfluidity in its core.
   However, the implications of neutron \superfluidp{} superfluidity for the cooling of neutron stars after an episode of mass accretion from a stellar companion in a low-mass X-ray binary has received little attention.
   }
   {We investigate the role of neutron \superfluidp{} superfluidity on the cooling of transiently accreting neutron stars after an outburst and compare the evolution of the effective surface temperature with observations of four different sources: \mxb, \ks, \igr{} and \xte{}.
   } 
   {Fully general relativistic simulations of neutron star cooling are performed using the code \nscool. We consider seven neutron \superfluidp{} pairing gaps, six of them based on microscopic calculations with three different realistic nucleon-nucleon potentials and different many-body approximations, and one directly inferred from the rapid cooling of Cassiopeia A.}
   {We find that the presence of neutron \superfluidp{} superfluidity provides a very modest improvement to the fit of the cooling data.
   The $\chi^2$ values with and without this neutron pairing channel differ by less than $10\%$. With the four sources studied, no neutron \superfluidp{} pairing gap can be singled out.
   We find modest evidence in favor of gaps corresponding to $T_{c,n,\mathrm{max}} \approx 0.4-0.6$ GK, compatible with the constraints from the cooling of Cassiopeia A.
   }
   {The presence or absence of neutron \superfluidp{} superfluidity in neutron stars cannot be ruled out from current cooling data of transient X-ray sources.
   However, we raise the possibility that the \superfluidp{} pairing gap could be more tightly constrained by further observations of \igr{} or \xte.}


   \maketitle
   \nolinenumbers

\section{Introduction}\label{sect:intro}

The observed rapid cooling of the isolated neutron star in the supernova remnant of Cassiopeia A \citep{HoHeinke_2009, HeinkeHo_2010, Posselt_2018, Wijngaarden_2019, Ho_2021, Posselt_2022,Zhao_2025} is generally attributed to the onset of a phase transition to neutron \superfluidp{} superfluidity in the dense core \citep{Page_2011PhRvL.106h1101P, Shternin_2011MNRAS.412L.108S} due to enhanced neutrino emission from neutron pair breaking and formation (PBF) \citep{1999PhyU...42..737Y}. The modeling of this astrophysical scenario constrains the maximum critical temperature for neutron \superfluidp{} pairing $T_{c,\mathrm{max}}$ to lie in the range of values between 0.55 to 0.9 GK in agreement with results from posterior works \citep[e.g.][]{Ho_2015, Shternin_2021, Shternin_2023}.
The aforementioned theoretical interpretation is not unique. Particularly, in later years some authors have called out for alternative interpretations. For instance, \citet{Negreiros_2013, Taranto_2016} argued that the direct Urca process in the core might be operating in this source; more recently \citet{Leinson_2022} argued in favor of a hybrid scenario where both direct Urca and neutron \superfluidp{} pairing are at operation in the core, whereas \citet{Potekhin_2026JHEAp..4900441P} proposed that either the direct or the modified versions of these processes might be the sole responsible for the rapid cooling of Cassiopeia A. These explanations are plausible given that both PBF and Urca processes can produce a high neutrino emissivity. It must be noted, however, that neutron \superfluidp{} also leads to a suppression of the associated neutron specific heat as well as the neutrino emissivity of the direct Urca process.

From a microphysical point of view, the existence of neutron \superfluidp{} superfluidity remains uncertain. 
The associated critical temperature and pairing gap depend sensitively on the many-body treatment \citep[see, for instance][]{Ding_2016, Dong_2016, Sedrakian_2019, Krotscheck_2024}. In contrast, neutron \superfluids{} superfluidity is well established and calculations of its properties based on different approaches are in fairly good agreement  \citep[see, for instance][]{Cao_2006,Gandolfi_2022,Drissi_2022}. The uncertainties in the neutron \superfluidp{} gaps on the cooling of Cassiopeia A were investigated by \citet{Ho_2015, Leinson_2022,Potekhin_2026JHEAp..4900441P}. Neutron \superfluidp{} superfluidity was independently probed by \citet{Kantor_2020} through the $r$-mode instability in Low-Mass X-ray binaries although empirical gaps were adopted rather than microscopic ones.

With respect to these systems, in particular those having transient accretion episodes, the observed cooling of neutron stars during quiescence provide a unique opportunity for expanding our knowledge of the properties of their crust and the amount of heat deposited during accretion. 
Little has been said regarding neutron superfluidity in the \superfluidp{} channel unlike the \superfluids{} channel \citep[see for instance][]{Turlione_2015A&A...577A...5T,Abbott_2017_GW,Allard_2024EPJA}. This stems from the fact that up to about $10^{3}$ days after the end of an outburst episode, the cooling of the neutron star is controlled by the heat transport in their crust. Still, the effects of neutron \superfluidp{} superfluidity was considered by \citet{Ootes_2019A&A...630A..95O, Fortin_2018MNRAS.475.5010F}, although these studies were limited to  understanding the long-term evolution of transiently accreting neutron stars over a period of about $10^5$ years.  \citet{Ootes_2019A&A...630A..95O} discussed only the general physical implications of neutron \superfluidp{} superfluidity, as they performed a sensitivity study of the many ingredients involved in cooling simulations. However, these authors employed a Gaussian-like parametrization of the neutron \superfluidp{} pairing gap, which might close at higher densities than predicted by microscopic calculations. In contrast, \citet{Fortin_2018MNRAS.475.5010F} actually performed a comparison of theoretical expectations against observations albeit including only two neutron \superfluidp{} pairing gaps, with one of them mimicking the presence of polarization effects through a multiplicative factor. Their conclusion regarding  $T_{c,\mathrm{max}}$ is consistent with constraints inferred from the rapid cooling of Cassiopeia A.
On a shorter time scale, the cooling of transiently accreting neutron stars $10^{3}$ days after the end of an outburst is dominated by the core physics and could therefore also place some 
further constraint on neutron \superfluidp{} superfluidity. For some sources 
such as \ks{}, \citet{Turlione_2015A&A...577A...5T} fitted the data by fine-tuning
the neutron \superfluids{} pairing gap extended to the core region, precisely where \superfluidp{} pairing is expected to operate.

In this work, we revisit the cooling of transiently accreting neutron stars during quiescence allowing for neutron \superfluidp{} superfluidity in their core, and compare our predictions with observations. We consider both realistic pairing gaps based on microscopic calculations and a gap directly inferred from the cooling of Cassiopeia A.

In Section~\ref{sect:methods}, we describe our numerical setup. In Section~\ref{sect:results}, we discuss the results for the transient sources. Discussion are presented in Section~\ref{sect:discussion} and conclusions are stated in Section~\ref{sect:conclusions}.
\section{Methodology}\label{sect:methods}

To simulate the thermal evolution of neutron stars, we employ the latest version of the code \nscool{} \citep{2016ascl.soft09009P, Page_2022, Nava-Callejas_2025}. This code solves the general-relativistic evolution equations of temperature and luminosity throughout the crust and the core of the neutron star.

For the core, we adopt the BSk24 equation of state (EOS) \citep{Pearson_2018}. The tabulated versions we employ is available in the CompOSE repository\footnote{\url{https://compose.obspm.fr}} \citep{Compose_1_Typel2015, Compose_2_Oertel2017, Compose_3_team}. To ensure the neutron star cooling is not affected by the fast direct Urca mechanism, we choose a star with a mass of $1.4 M_{\odot}$ (for reference, the critical neutron star mass for the onset of direct Urca with the BSk24 EOS is $\sim 1.6 M_{\odot}$, see \citet{Pearson_2018}). This value is also consistent with the one employed in the spectral fittings for the sources under study \citep[e.g.][]{Merritt_2016ApJ...833..186M, Parikh2019, Parikh_2020}.
For the envelope boundary condition of the neutron star, we use the common $T_b - T_\mathrm{eff}$ prescription which parametrizes the presence of light elements at the surface via a column depth $y_{L}$ \citep[e.g.][]{2009ApJ...698.1020B, Parikh2019}.
As is usual in the field, to parametrize the thermal conductivity of the neutron star crust we employ an impurity parameter $Q_\mathrm{imp}$, assigning one value for the inner crust and another for the outer crust. Following current estimations for this parameter, based on calculations of ashes from the rp-process and superbursts \citep{Lau_2018, 2025PhRvL.135p2701J},  $Q_\mathrm{imp}$ is bounded between $10^{-3}$ to $250$.

\subsection{Pairing gaps}\label{subsec:methods_gaps}

\begin{figure}
    \centering
    \includegraphics[width=1.0\linewidth]{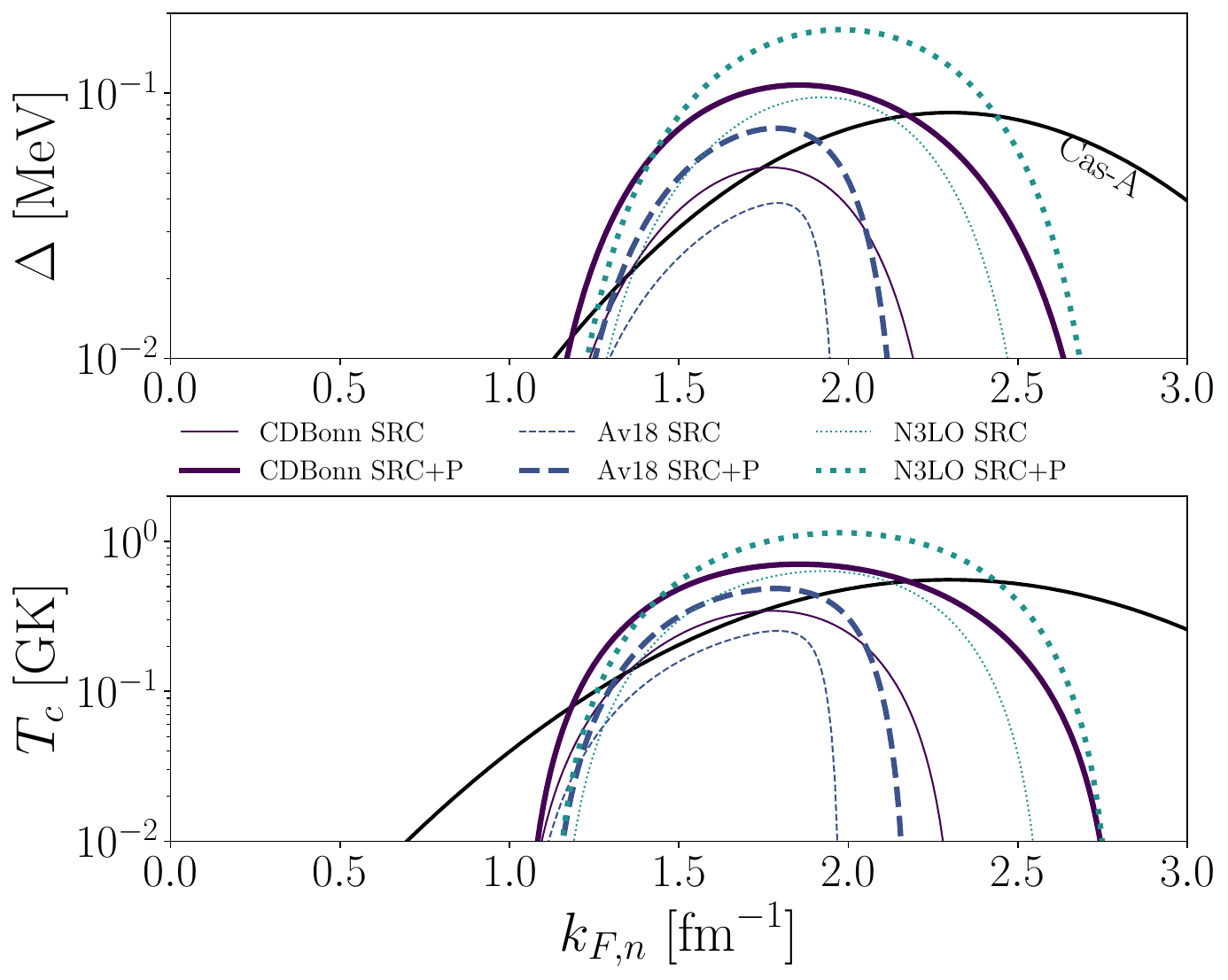}
\caption{Upper panel: Neutron \superfluidp{} pairing gaps (in MeV) employed in the present work as a function of the neutron Fermi wavenumber (in fm$^{-1}$). Lower panel: corresponding critical temperature $T_{c}$ (in units of GK) for each of the gaps from the upper panel.}
\label{fig:pairing_gaps}
\end{figure}

For the neutron \superfluids{} pairing gap, we adopt the quantum Monte Carlo calculations of \citet{Gandolfi_2022}, labeled QMCG. For the proton \superfluids{} pairing gap, we adopt the classical results from \citet{Chen_1993}, labeled as CCDK.
For the neutron \superfluidp{} pairing gap, we consider seven options, displayed as functions of the neutron Fermi wavenumber $k_{F,n}$ in Fig.~\ref{fig:pairing_gaps}. The first corresponds to the gap empirically adjusted to reproduce the rapid cooling of Cassiopeia A, as reported in \citet{Page_2011PhRvL.106h1101P}, and accordingly labeled as Cas-A. The remaining six gaps are taken from the microscopic calculations by \citet{Ding_2016}. These authors computed the pairing gap considering three different realistic nucleon-nucleon potentials: CDBonn, Av18 and N3LO. In their basic form, all their computed gaps include short-range correlations (SRC). The choice of the potential changes the size of the gap. Ordered from smallest to largest, we have CDBonn, Av18 and N3LO. The second family of gaps, denoted as SRC+P, corresponds to the inclusion of long-range correlations (LRC), which have the potential-independent effect of further enhancing the size of the gap. The absolute maximum of all the gaps, however, does not exceed $\sim 0.2$ MeV. All the gaps calculated by \citet{Ding_2016} vanish for $k_{F,n}$ beyond 2.7 fm$^{-1}$. In contrast, the Cas-A gap closes at a much larger wavenumber, $\sim$3.5 fm$^{-1}$. At low density, however, all gaps close at around 1 fm$^{-1}$.
The gaps considered here are representative to those employed in former cooling studies, as well as those based on recent microscopic calculations. An explicit comparison can be found in Appendix~\ref{appendix:gaps_comparison}.


\subsection{Observational sources and fitting method}\label{subsec:methods_fittings}

\begin{table}
\scriptsize
\begin{center}
\caption{Compilation of the observational data for the transient X-ray sources considered in this work.}
\begin{tabular}{|p{2.0cm}|p{0.55cm}|p{1.5cm}|p{2.0cm}|}
\hline
Source & $t_{\mathrm{acc}}$ & $\langle\dot{M}\rangle_\mathrm{obs}$ & Refs.\\
& [yr] & $[10^{18}\ \mathrm{g\ s}^{-1}]$ & \\
\hline
%
\mxb (OI) \newline{(\mxbi)} & 2.500  & $0.120$  & \cite{Parikh2019}\\
\hline
\ks\newline(\ksbrief)    & 12.50  & $0.15$ & \cite{Merritt_2016ApJ...833..186M}; \citet{Ootes_2016}\\
\hline
\igrS\newline(\igrbrief)  & 0.170  & $0.19$ & \cite{Degenaar_2011, Ootes_2019}\\
\hline
\xte\newline(\xtebrief)    & 1.600  & 1.1 & \cite{Parikh_2020}\\
\hline
\end{tabular}
\label{tab:observational_data}
\end{center}
\end{table}

In this work, we focus on four transient sources: \mxb, \ks{}, \igr{} and \xte, hereafter \mxbbrief{}, \ksbrief{}, \igrbrief{} and \xtebrief{}, respectively. Each outburst episode is modeled as a period of constant mass accretion at a rate $\dot{M}$, lasting for $t_\mathrm{acc}$. The first two sources were chosen because, among the known cooling transient sources, they display the lowest effective temperature, $< 70$ eV, $10^{4}$ days after the end of their respective outbursts. The latter two were chosen for precisely the opposite reason, i.e., for displaying a high effective temperature at this period, $\gtrsim 80$ eV. The accretion time and the average mass accretion rate are listed in Table~\ref{tab:observational_data}. In the same table we include the references from which these data are taken, as well as the cooling data.

To fit the data, we use the Markov Chain Monte Carlo (MCMC) method \citep[see, e.g.][]{Lin_2018, Ootes_2019, Degenaar_2021, Page_2022}. Here we aim to explain the data using five free parameters: the core's redshifted temperature \tcore{}, the maximum column depth of light elements $y_{L}$, two impurity parameters corresponding to the outer and inner crust, $Q_\mathrm{imp,1}$ and $Q_{\mathrm{imp,2}}$, and the redshifted envelope-crust boundary \tb.

We consider 8 models for each source. The first one, a simulation without neutron \superfluidp{} pairing, is taken as the baseline model. The remaining seven simulations correspond to the gaps described in the previous subsection.
\section{Results}\label{sect:results}

\begin{table*}
\scriptsize
\begin{center}
\caption{Mean values of cooling parameters and their $90\%$ confidence interval for each source and choice of \superfluidp{} neutron pairing gap.}
\begin{tabular}{|c|c|c|c|c|c|c|c|}
\hline
Source & \superfluidp{} gap & \tcore & $\log_{10}y_{L}$ & $Q_{\mathrm{imp,1}}$ & $Q_{\mathrm{imp,2}}$ & \tb & $\chi^{2}(\mathrm{d.o.f.})$ \\
& & $[10^{7}\ \mathrm{K}]$ & $[10^{8}\ \mathrm{g\ cm}^{-2}]$ & & & $[10^{8}\ \mathrm{K}]$ & \\
\hline
\mxbi & No  & $3.37^{+1.11}_{-0.61}$ & $8.22^{+1.56}_{-1.94}$ & $2.18^{+28.48}_{-2.17}$ & $3.35^{+3.30}_{-3.34}$ & $5.29^{+0.65}_{-0.70}$ & 6.24 (2)  \\
& Cas-A & $3.12^{+1.17}_{-0.77}$ & $8.21^{+1.59}_{-1.68}$ & $14.98^{+15.50}_{-14.96}$ & $0.68^{+5.12}_{-0.67}$ & $5.22^{+0.57}_{-0.70}$ & 6.25 (2)  \\
& \dingi & $3.20^{+1.07}_{-0.67}$ & $8.48^{+1.38}_{-2.15}$ & $5.76^{+25.13}_{-5.76}$ & $2.92^{+3.45}_{-2.92}$ & $5.16^{+0.69}_{-0.62}$ & 6.15 (2) \\
& \dingii & $3.02^{+1.10}_{-0.59}$ & $8.34^{+1.52}_{-1.88}$ & $7.25^{+24.33}_{-7.25}$ & $2.90^{+3.37}_{-2.90}$ & $5.22^{+0.65}_{-0.75}$ & 6.22 (2) \\
& \dingiii & $3.33^{+0.98}_{-0.62}$ & $8.55^{+1.31}_{-2.01}$ & $3.43^{+23.65}_{-3.42}$ & $3.11^{+2.83}_{-3.10}$ & $5.16^{+0.74}_{-0.64}$ & 6.21 (2) \\
& \dingiv  & $3.18^{+1.02}_{-0.61}$ & $8.39^{+1.46}_{-1.71}$ & $1.20^{+26.55}_{-1.20}$ & $3.53^{+3.03}_{-3.52}$ & $5.22^{+0.69}_{-0.66}$ & 6.19 (2) \\
& \dingv & $2.99^{+1.18}_{-0.69}$ & $8.46^{+1.37}_{-2.13}$ & $7.02^{+26.31}_{-7.02}$ & $2.77^{+3.12}_{-2.76}$ & $5.19^{+0.69}_{-0.66}$ & 6.30 (2) \\
& \dingvi & $3.17^{+1.02}_{-0.78}$ & $8.13^{+1.68}_{-1.66}$ & $14.30^{+17.28}_{-14.29}$ & $1.21^{+4.56}_{-1.21}$ & $5.24^{+0.64}_{-0.79}$ & 6.19 (2) \\

\hline

\ksbrief & No & $7.72^{+0.74}_{-1.17}$ & $6.59^{+0.89}_{-0.53}$ & $0.31^{+16.39}_{-0.30}$ & $0.11^{+2.04}_{-0.11}$ & $5.27^{+0.30}_{-0.55}$ & 2.98 (3) \\
& Cas-A & $7.53^{+0.75}_{-2.00}$ & $6.64^{+1.56}_{-0.58}$ & $0.72^{+18.18}_{-0.72}$ & $0.12^{+2.17}_{-0.12}$ & $5.25^{+0.34}_{-0.66}$ & 2.92 (3)  \\
& \dingi & $7.74^{+0.66}_{-1.37}$ & $6.55^{+1.13}_{-0.49}$ & $0.26^{+11.60}_{-0.26}$ & $0.09^{+2.21}_{-0.09}$ & $5.32^{+0.29}_{-0.55}$ & 3.00 (3)  \\
& \dingii & $7.05^{+0.92}_{-1.94}$ & $6.65^{+1.47}_{-0.60}$ & $0.61^{+18.27}_{-0.61}$ & $0.16^{+2.35}_{-0.16}$ & $5.30^{+0.31}_{-0.65}$ & 2.81 (3)  \\
& \dingiii & $7.88^{+0.57}_{-1.57}$ & $6.48^{+1.36}_{-0.43}$ & $0.18^{+12.84}_{-0.18}$ & $0.09^{+1.85}_{-0.08}$ & $5.30^{+0.28}_{-0.54}$ & 2.93 (3) \\
& \dingiv & $7.61^{+0.76}_{-1.37}$ & $6.56^{+1.17}_{-0.52}$ & $0.25^{+17.45}_{-0.25}$ & $0.21^{+1.74}_{-0.21}$ & $5.29^{+0.33}_{-0.58}$ & 2.98 (3) \\
& \dingv & $7.28^{+0.87}_{-1.90}$ & $6.63^{+1.42}_{-0.57}$ & $0.46^{+14.26}_{-0.45}$ & $0.11^{+2.48}_{-0.11}$ & $5.30^{+0.30}_{-0.68}$ & 2.85 (3) \\
& \dingvi & $6.97^{+0.84}_{-2.20}$ & $6.61^{+1.81}_{-0.56}$ & $0.17^{+19.51}_{-0.17}$ & $0.27^{+2.81}_{-0.27}$ & $5.29^{+0.32}_{-0.99}$ & 2.76 (3) \\

\hline

\igrbrief & No & $11.17^{+3.12}_{-1.85}$ & $8.47^{+1.41}_{-1.85}$ & $90.42^{+144.23}_{-90.41}$ & $0.37^{+97.76}_{-0.37}$ & $5.20^{+1.45}_{-1.11}$ & 11.83 (4) \\
& Cas-A & $11.10^{+2.62}_{-1.83}$ & $8.76^{+1.13}_{-1.79}$ & $104.05^{+134.72}_{-104.05}$ & $0.42^{+80.89}_{-0.41}$ & $5.09^{+1.58}_{-1.21}$ & 11.98 (4) \\
& \dingi & $11.29^{+3.17}_{-1.99}$ & $8.48^{+1.41}_{-2.25}$ & $63.07^{+169.25}_{-63.06}$ & $0.32^{+58.03}_{-0.32}$ & $5.47^{+1.25}_{-1.31}$ & 12.18 (4) \\
& \dingii & $11.36^{+2.95}_{-2.10}$ & $8.54^{+1.36}_{-2.05}$ & $107.79^{+131.93}_{-107.78}$ & $0.47^{+67.02}_{-0.47}$ & $5.14^{+1.36}_{-1.28}$ & 11.81 (4) \\
& \dingiii & $11.03^{+3.08}_{-1.92}$ & $8.69^{+1.23}_{-2.55}$ & $27.52^{+200.97}_{-27.52}$ & $0.79^{+84.11}_{-0.79}$ & $5.77^{+0.97}_{-1.71}$ & 11.76 (4) \\
& \dingiv  & $12.27^{+2.39}_{-2.89}$ & $7.67^{+2.07}_{-1.59}$ & $75.76^{+160.85}_{-75.76}$ & $1.32^{+94.38}_{-1.32}$ & $5.47^{+1.34}_{-1.22}$ & 11.84 (4) \\
& \dingv & $11.32^{+3.10}_{-2.06}$ & $8.58^{+1.32}_{-2.01}$ & $107.38^{+130.09}_{-107.37}$ & $0.27^{+35.32}_{-0.27}$ & $5.13^{+1.49}_{-1.12}$ & 12.00 (4) \\
& \dingvi & $11.14^{+3.48}_{-1.93}$ & $8.48^{+1.41}_{-2.23}$ & $73.68^{+160.56}_{-73.67}$ & $0.52^{+73.49}_{-0.52}$ & $5.32^{+1.34}_{-1.21}$ & 12.17 (4) \\

\hline

\xtebrief & No & $24.23^{+3.66}_{-3.57}$ & $8.51^{+1.38}_{-2.21}$ & $210.53^{+37.57}_{-209.69}$ & $0.09^{+29.24}_{-0.09}$ & $4.97^{+0.68}_{-0.95}$ & 10.36 (8) \\
& Cas-A & $25.11^{+2.91}_{-3.88}$ & $8.05^{+1.78}_{-1.76}$ & $197.25^{+50.23}_{-197.16}$ & $0.33^{+39.67}_{-0.33}$ & $5.15^{+0.59}_{-0.97}$ & 10.18 (8)  \\
& \dingi & $24.33^{+3.41}_{-3.28}$ & $8.39^{+1.48}_{-2.01}$ & $210.88^{+37.76}_{-202.89}$ & $0.12^{+26.99}_{-0.12}$ & $5.02^{+0.65}_{-0.88}$ & 10.21 (8) \\
& \dingii & $25.66^{+3.06}_{-4.26}$ & $8.04^{+1.78}_{-1.85}$ & $196.96^{+50.76}_{-196.73}$ & $0.17^{+11.45}_{-0.17}$ & $5.12^{+0.61}_{-1.00}$ & 10.29 (8) \\
& \dingiii & $23.45^{+4.00}_{-2.56}$ & $8.81^{+1.04}_{-1.91}$ & $209.74^{+38.34}_{-206.62}$ & $0.19^{+22.38}_{-0.19}$ & $4.97^{+0.72}_{-0.96}$ & 10.43 (8) \\
& \dingiv & $24.46^{+3.75}_{-3.55}$ & $8.41^{+1.44}_{-2.00}$ & $190.80^{+56.36}_{-190.80}$ & $0.16^{+24.76}_{-0.16}$ & $5.14^{+0.71}_{-0.94}$ & 10.37 (8) \\
& \dingv & $24.04^{+4.08}_{-3.28}$ & $8.60^{+1.29}_{-2.36}$ & $204.63^{+42.64}_{-187.15}$ & $0.14^{+35.09}_{-0.14}$ & $5.00^{+0.68}_{-0.97}$ & 10.17 (8) \\
& \dingvi & $25.91^{+3.26}_{-4.69}$ & $7.96^{+1.84}_{-1.84}$ & $197.73^{+49.53}_{-197.73}$ & $0.14^{+25.11}_{-0.14}$ & $5.12^{+0.66}_{-0.99}$ & 10.18 (8) \\

\hline
\end{tabular}
\tablefoot{From left to right: core's temperature \tcore, the maximum column depth at which light elements are located, $\log_{10}y_{L}$; $Q_\mathrm{imp,1}$ and $Q_\mathrm{imp,2}$ correspond to the impurities in the outer and inner crust, and finally \tb{} denotes the redshifted temperature at the envelope-crust boundary. The $\chi^2$ for the best fitting model in each case is reported.}
\label{tab:results_all_sources}
\end{center}
\end{table*}

\begin{figure*}
    \sidecaption
    \includegraphics[width=0.7\linewidth]{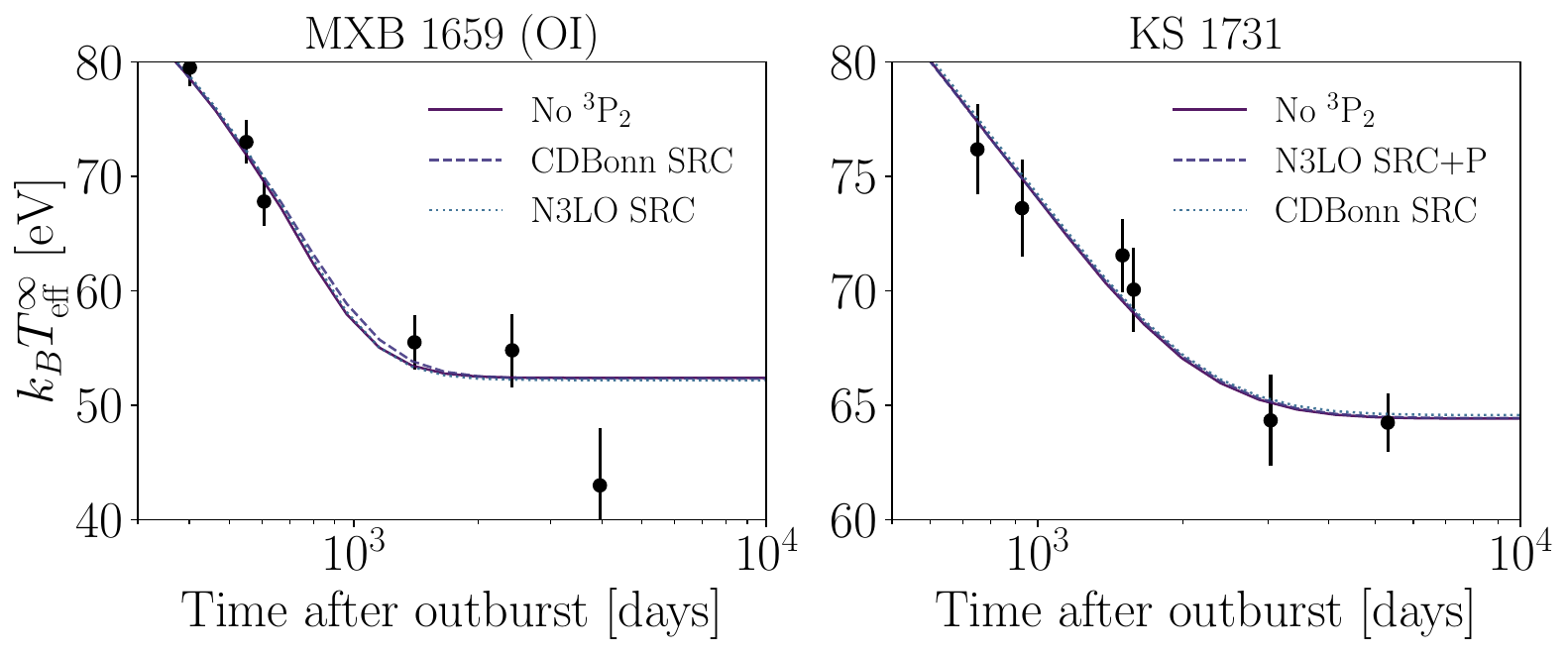}
\caption{Best fitting cooling curves after an outburst of \mxb{} and \ksbrief. Continuous line corresponds to the model without neutron \superfluidp{} pairing, dashed line to the model with the lowest $\chi^2$ and the dotted curve to the model with the highest $\chi^2$ (see Table~\ref{tab:results_all_sources} for the numerical values).}
\label{fig:best_fitting_curves_1}
\end{figure*}

\begin{figure*}
    \sidecaption
    \includegraphics[width=0.7\linewidth]{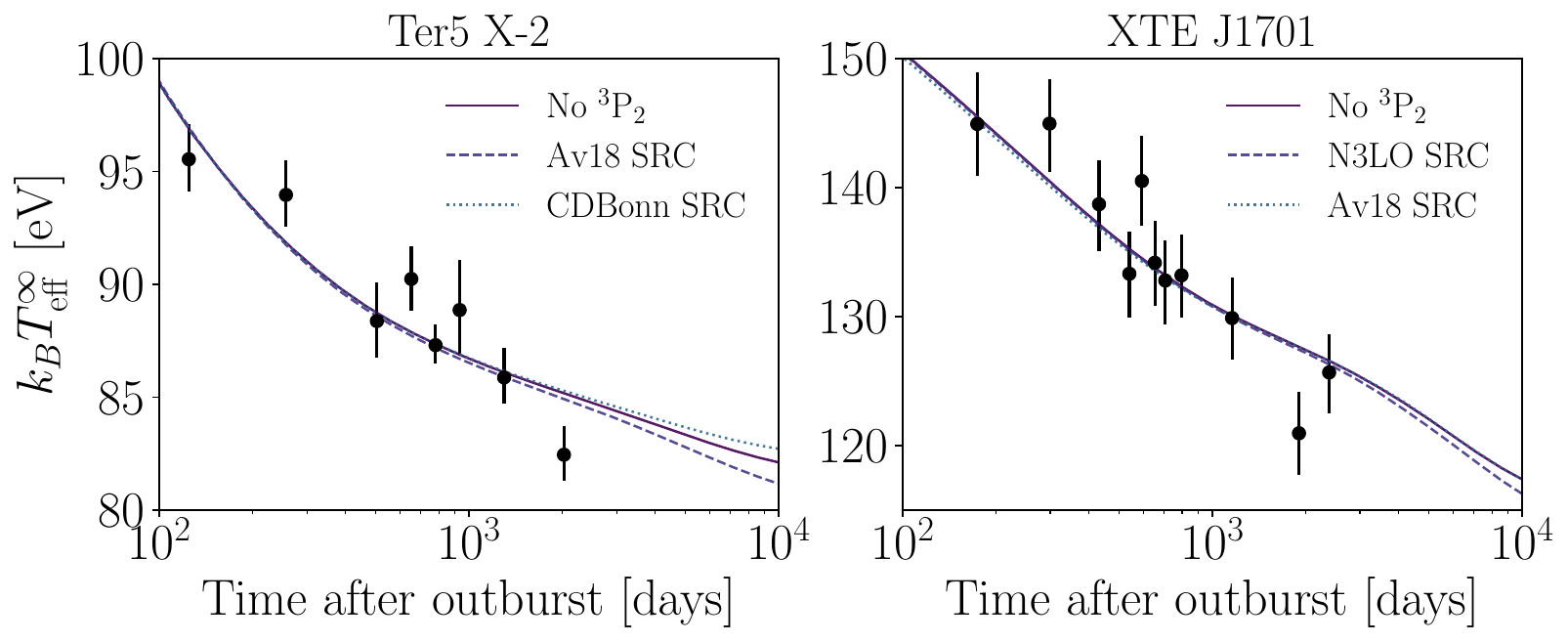}
\caption{Same as Fig.~\ref{fig:best_fitting_curves_1}, now for \igrbrief{} and \xtebrief.}
\label{fig:best_fitting_curves_2}
\end{figure*}

In Table~\ref{tab:results_all_sources}, we report the mean values of the model parameters and their 90\% confidence intervals for all scenarios, together with the corresponding $\chi^2$ values and number of degrees of freedom.
Figures~\ref{fig:best_fitting_curves_1} and~\ref{fig:best_fitting_curves_2} show three fitted cooling curves for each source. The solid line represents the baseline model without neutron \superfluidp{} pairing, while the dashed and dotted lines represent the models with neutron \superfluidp{} pairing that yield the lowest and highest $\chi^2$, respectively.
Overall, we find a similar level of agreement between the baseline model and the models including neutron \superfluidp{} pairing. For each source, at least one of the pairing gaps yields a lower $\chi^2$ than the baseline model. The extent of this improvement depends on both the source and the assumed pairing gap. For instance, the \dingv{} and \dingvi{} gaps yield substantially lower $\chi^2$ values than the baseline model for \ksbrief{} and \xtebrief{}, whereas the opposite trend is found for \igrbrief{}.
\section{Discussion}\label{sect:discussion}

\subsection{On the physical properties of the four sources}\label{subsec:discuss_1}

\begin{figure}
    \includegraphics[width=1.0\linewidth]{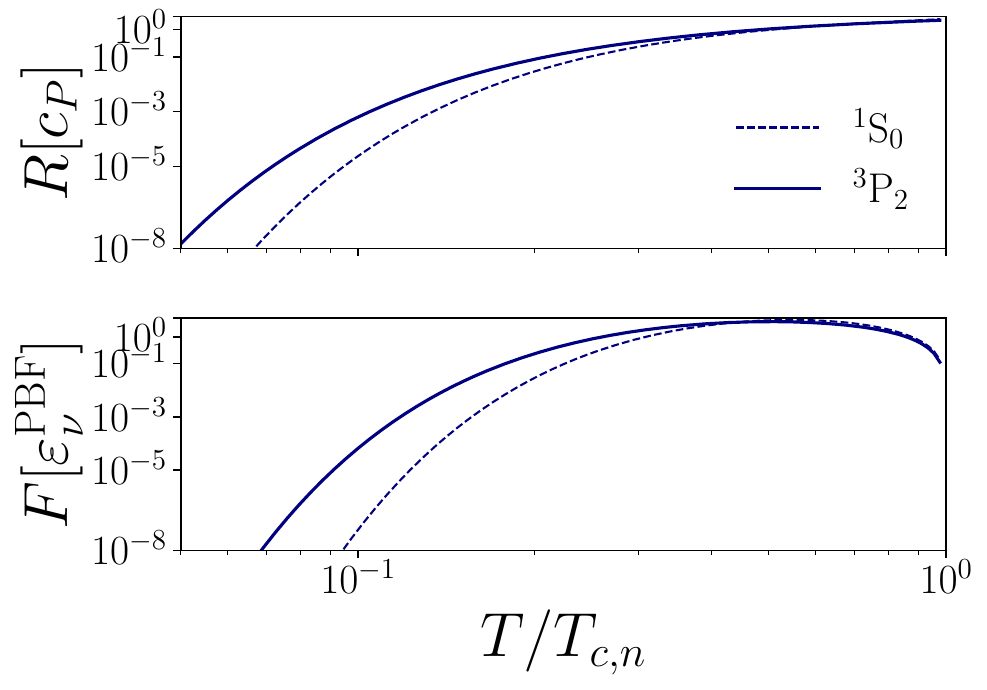}
\caption{Upper panel: suppression factor of the neutron specific heat as a function of $T/T_{c,n}$ for both \superfluids{} and \superfluidp{} pairing. Lower panel: neutrino emissivity factor associated with the \superfluids{} and \superfluidp{} pair breaking and formation (PBF) mechanism as a function of $T/T_{c,n}$.}
\label{fig:neutrino_factors}
\end{figure}

\begin{figure*}
    \centering
    \includegraphics[width=0.4\linewidth]{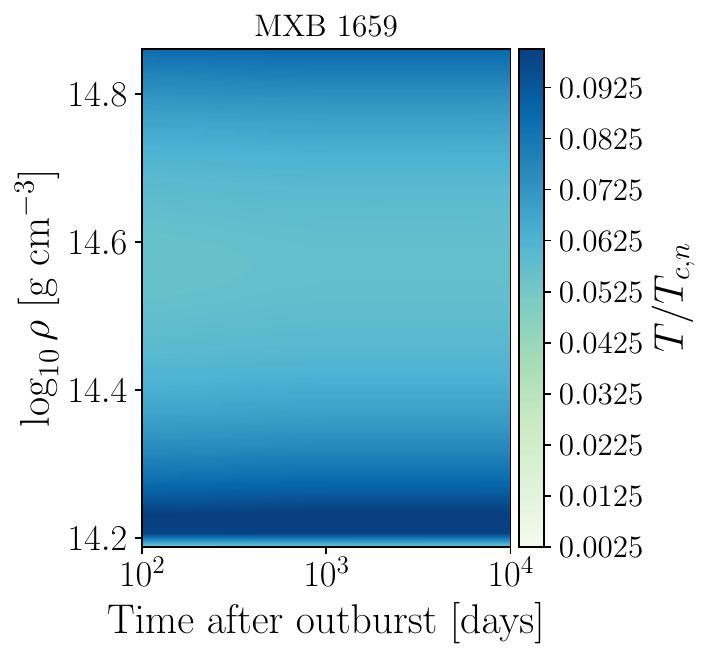}
    \includegraphics[width=0.4\linewidth]{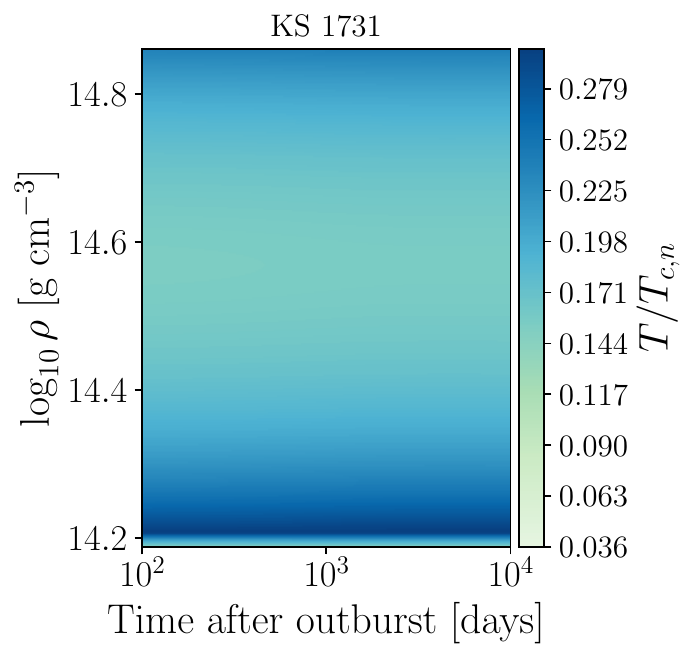}
\caption{$T/T_{c,n}$ contours for the coldest sources, \mxbbrief{} and \ksbrief, computed with the pairing gap \dingii.}
\label{fig:t_tcn_all_1}
\end{figure*}

\begin{figure*}
    \centering
    \includegraphics[width=0.4\linewidth]{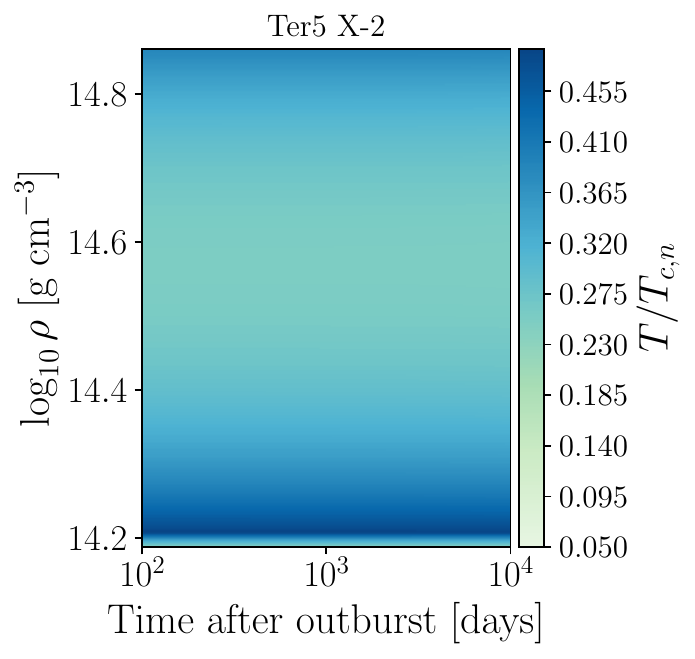}
    \includegraphics[width=0.4\linewidth]{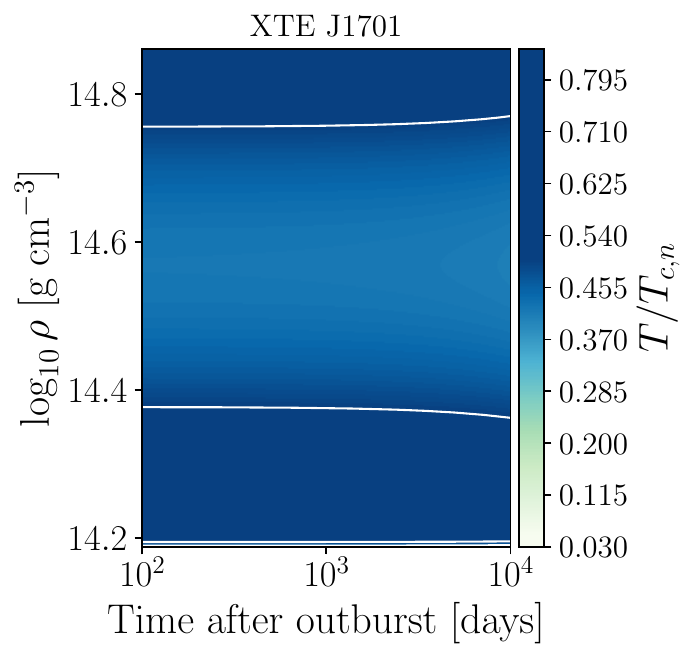}
\caption{Same as Fig.~\ref{fig:t_tcn_all_1}, now for \igrbrief{} and \xtebrief.}
\label{fig:t_tcn_all_2}
\end{figure*}

Since neutron stars are assumed to be not massive enough as to trigger the direct Urca process, the presence of neutron \superfluidp{} pairing has a major effect for the cooling of the core through two effects: first, the suppression of the neutron specific heat and second the introduction of a source of neutrino emission. Both depend primarily on the ratio $T/T_{c,n}$, where $T$ is the local temperature and $T_{c,n}$ is the critical temperature for neutron superfluidity. In order to make the discussion as self-contained as possible, in Fig.~\ref{fig:neutrino_factors} we display the suppression factor of the neutron specific heat and the neutrino emissivity factor associated with each \superfluids{} and \superfluidp{} PBF mechanism (see Appendix~\ref{appendix:3p2_r_f}). These curves were computed employing the fits from \citet{1999PhyU...42..737Y}. Further details can be found in Appendix~\ref{appendix:3p2_r_f}. For $T/T_{c,n}\ll 0.1$, the neutron specific heat is almost completely suppressed, while PBF neutrino emission is exponentially small. In this regime, the presence of neutron \superfluidp{} superfluidity thus mainly manifests as a suppression of the neutron specific heat. For $T/T_{c,n}\sim 0.1-1$, the neutron specific heat is only moderately suppressed while PBF emission becomes significant, first from the neutron \superfluids{} channel (because of the higher critical temperature) and, as the star cools down, from the neutron \superfluidp{} channel. In this second regime, we then expect that the presence of neutron \superfluidp{} superfluidity influences the cooling throughout the fast neutrino emission and only moderately the suppression of the neutron specific heat.
With this framework in mind, let us now turn to the inferred physical parameters. We find that \tcore{} is relatively well constrained for each outburst and varies only modestly with the assumed neutron \superfluidp{} pairing prescription. For \mxbi{}, we obtain \tcore{} $\sim 3\times10^{7}$ K; for \ksbrief{}, $\sim 7.5\times10^{7}$ K; for \igrbrief{}, $\sim 11\times10^{7}$ K; and for \xtebrief{}, $\sim 23\times10^{7}$ K. In terms of local temperatures, these values correspond to\footnote{The appropriate redshift factor for the considered gravitational mass and radius is 0.80.} $3.8\times 10^{7}$ K, $9.4\times 10^{7}$ K, $13.8\times 10^{7}$ K and $28.8\times 10^{7}$ K. 
For \tb{} and $y_{L}$ we also find relatively well constrained values, whereas the preferred values of $Q_\mathrm{imp,1}$ and $Q_\mathrm{imp,2}$ depend more strongly on the adopted pairing gap.
Taking 0.5 GK as a characteristic scale for $T_{c,n,\mathrm{max}}$, the inferred local temperatures correspond approximately to $T/T_{c,n}\sim 0.08$ for \mxbbrief{}, $\sim 0.20$ for \ksbrief{}, $\sim 0.28$ for \igrbrief{}, and $\sim 0.58$ for \xtebrief{}. These estimates show that all four sources have core temperatures below the characteristic critical temperature, but they probe different regimes. \mxbbrief{} and \ksbrief{} lie in the regime $T/T_{c,n}< 0.1$ where the suppression of the neutron specific heat drives the cooling of the star. In contrast, \igrbrief{} and \xtebrief{} lie in the intermediate regime $T/T_{c,n}\sim 0.1$--$0.6$ where the neutron specific heat is moderately suppressed and PBF neutrino emission can be significant. To further illustrate this point, in Figs.~\ref{fig:t_tcn_all_1} and~\ref{fig:t_tcn_all_2} we display the evolution of $T/T_{c,n}$ in the core for all four sources, considering the pairing gap \dingii. In these figures it is noticeable that a large portion of the core in all four sources is at $T/T_{c,n}< 1$.
The importance of neutron \superfluidp{} superfluidity is  also controlled by the density dependence of the pairing gap. 
This distinction is reflected in the fits. The two coolest sources, \mxbbrief{} and \ksbrief{}, exhibit a different behavior from the hottest sources, \igrbrief{} and \xtebrief{}. As can be seen in Figs.~\ref{fig:best_fitting_curves_1} and~\ref{fig:best_fitting_curves_2}, the latter sources have higher effective temperatures than the former, and the cooling curves suggest that these neutron stars have not thermally relaxed yet.
The level of improvement in $\chi^2$ obtained by including \superfluidp{} pairing also appears to correlate with the inferred core temperature, with larger improvements for the hotter sources. Given the small number of sources considered here, this association should be regarded as a tentative trend rather than evidence for a general correlation. It is, however, consistent with the expectation from Fig.~\ref{fig:neutrino_factors}: the cooler sources sit in a regime where the neutron specific heat is suppressed and the PBF neutrino emission is negligible, so the neutron superfluidity leaves little imprint on the cooling curve, whereas the hotter sources sample the intermediate regime where both effects vary with $T/T_{c,n}$.
As shown in Table~\ref{tab:results_all_sources}, the preferred values of $Q_\mathrm{imp,1}$ and $Q_\mathrm{imp,2}$ vary with the adopted \superfluidp{} pairing gap, thus introducing some degeneracies in the cooling models. This suggests that uncertainties in the \superfluidp{} pairing gap may propagate into the inferred crustal properties even when the resulting changes in the cooling data fit are modest.

\subsection{What do these sources tell us about the properties of the neutron \superfluidp{} pairing gap?}\label{subsec:discuss_2}

Table~\ref{tab:results_all_sources} reports the results for 32 models, four of which correspond to the baseline models without neutron \superfluidp{} superfluidity.
Of the remaining 28 models, 17 yield a lower $\chi^2$, ten yield a higher $\chi^2$, and one yields the same value. Thus, approximately $61\%$ of the models including neutron \superfluidp{} superfluidity provide a lower $\chi^2$ than their corresponding baseline models. While this fraction describes a trend in the behavior of the models considered here it does not, by itself, establish that neutron \superfluidp{} superfluidity is statistically favored. In particular, since no single \superfluidp{} pairing gap provides the best fit for all four sources. For instance, \mxbbrief{} favors the \dingi{} gap, whereas the cooling curve associated with the same gap gives the worst fit among the eight models considered for \igrbrief{}. Nevertheless, the results allow us to examine whether certain properties of the adopted gaps are associated with improvements in the $\chi^2$ for the four sources considered in this work.
\begin{itemize}
    \item \textbf{Maximum amplitude of the gap.}
    We can group the gaps according to their maximum value:  $\Delta_\mathrm{max}>10^{-1}$ MeV for the \dingvi{} gap, $\Delta_\mathrm{max}\sim 10^{-1}$ MeV for the Cas-A, \dingii{} and \dingv{} gaps, and $\Delta_\mathrm{max}<10^{-1}$ MeV for the \dingi{}, \dingiii{} and \dingiv{} gaps.
    Of the 17 model-source comparisons that improve on the baseline, three involve the first group, eight the second, and six the third. These counts do not reveal a clear monotonic dependence of fit quality on the maximum gap amplitude. They do, however, show that improvements occur for a range of amplitudes, including gaps whose maxima are below $0.1$ MeV. The amplitude alone is therefore insufficient to identify a single gap that consistently improves all the fits. However, our findings are rather consistent with the constraints on  $T_{c,n\mathrm{max}}$ inferred from the cooling of Cassiopeia A \citep{Page_2011PhRvL.106h1101P,Shternin_2011MNRAS.412L.108S, Shternin_2023}, 
    and $r$-mode instability in LMXB \citep{Kantor_2020}.
    \item \textbf{Gap closure at high density.}
    Figure~\ref{fig:pairing_gaps} shows that the adopted gaps can also be grouped according to the Fermi wavenumber at which they vanish: $k_{F,n}>3$ fm$^{-1}$ for the Cas-A gap, $k_{F,n}\geq 2.5$ fm$^{-1}$ for the \dingii{}, \dingv{} and \dingvi{} gaps, the \dingi{}, and $k_{F,n}<2.5$ fm$^{-1}$ for the \dingiii{} and \dingiv{} gaps.
    Among the 17 model-source comparisons that yield a lower $\chi^2$ than the baseline, two involve the Cas-A gap, nine involve gaps closing at $k_{F,n}\geq 2.5$ fm$^{-1}$, and six involve gaps closing at lower Fermi wavenumbers. This distribution suggests that, within the set of gaps considered here, improvements are more frequent for those that remain nonzero up to relatively high densities. In particular, all four fits obtained with the \dingii{} gap yield a lower $\chi^2$ than their corresponding baseline models. While this result is suggestive, it does not establish that a high-density closure is, in itself, responsible for the improvement. 
    \item \textbf{Fermi wavenumber at the maximum of the gap.}
    A further comparison concerns the location of the maximum. The \dingi{}, \dingiii{} and \dingiv{} gaps peak at $k_{F,n}\sim 1.77$--$1.79$ fm$^{-1}$, while the \dingii{}, \dingv{} and \dingvi{} gaps peak at somewhat higher Fermi momenta, between approximately $1.85$ and $1.98$ fm$^{-1}$. The Cas-A gap reaches its maximum at $k_{F,n}\approx 2.30$ fm$^{-1}$. Among the 17 improved model-source comparisons, six involve gaps in the first group, nine the second, and two the Cas-A gap. Thus, within the present sample, improvements are more frequent for gaps whose maxima lie between approximately $1.85$ and $1.98$ fm$^{-1}$ than for those peaking near $1.78$ fm$^{-1}$ or for the Cas-A gap. This possible association should be interpreted cautiously: the number of gaps is small, and the location of the maximum is correlated with other gap properties, including its amplitude and density range.
\end{itemize}
%


\subsection{On the relevance of neutron \superfluidp{} pairing in the forecast of the sources' long-term behavior}\label{subsec:discuss_3}

\begin{figure}
    \includegraphics[width=1.0\linewidth]{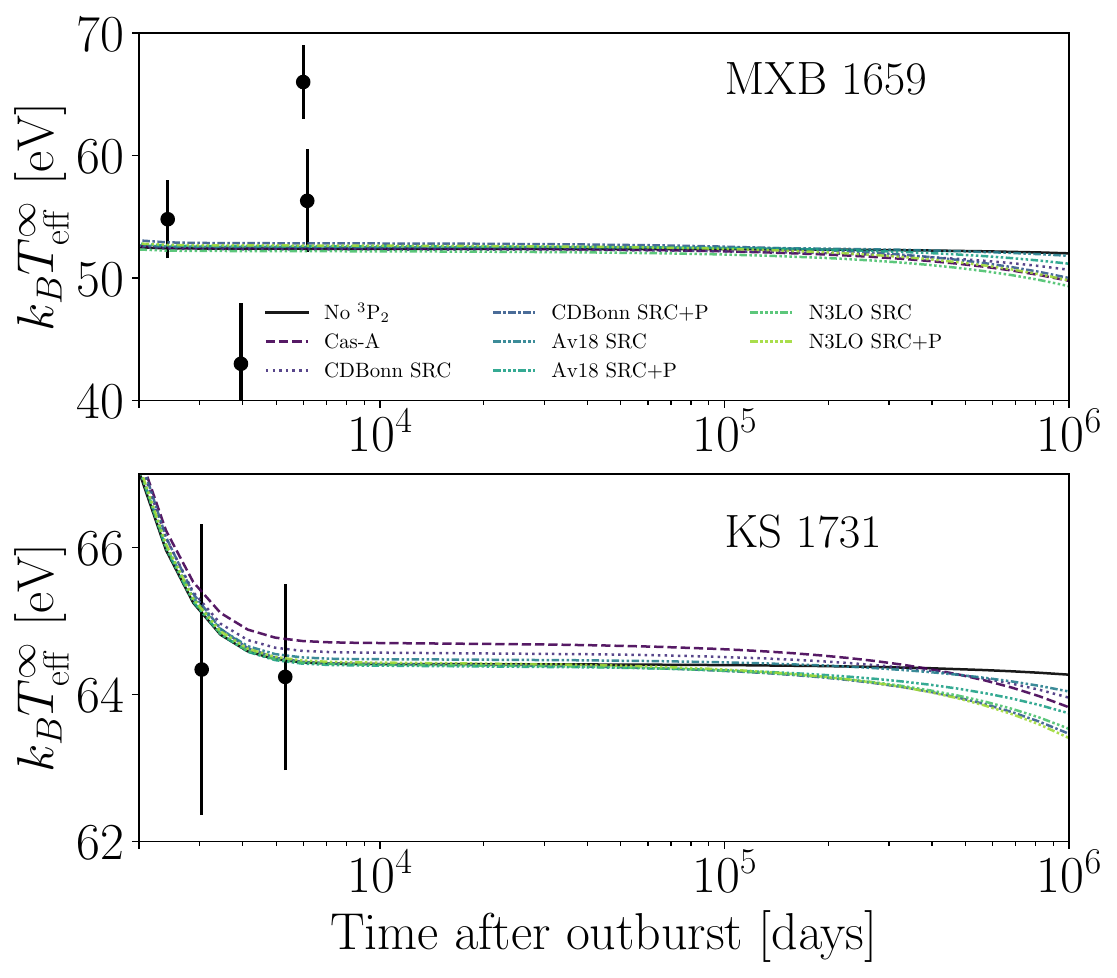}
\caption{Best-fitting cooling curves for all eight models reported in Table~\ref{tab:results_all_sources}, for the sources \mxbbrief{} and \ksbrief{}. The last two datapoints for \mxbbrief{}, which are not shown in Fig.~\ref{fig:best_fitting_curves_1}, correspond to the observed cooling after its second outburst in 2015. See the main text for further details.}
\label{fig:best_fitting_curves_1L}
\end{figure}

\begin{figure}
    \includegraphics[width=1.0\linewidth]{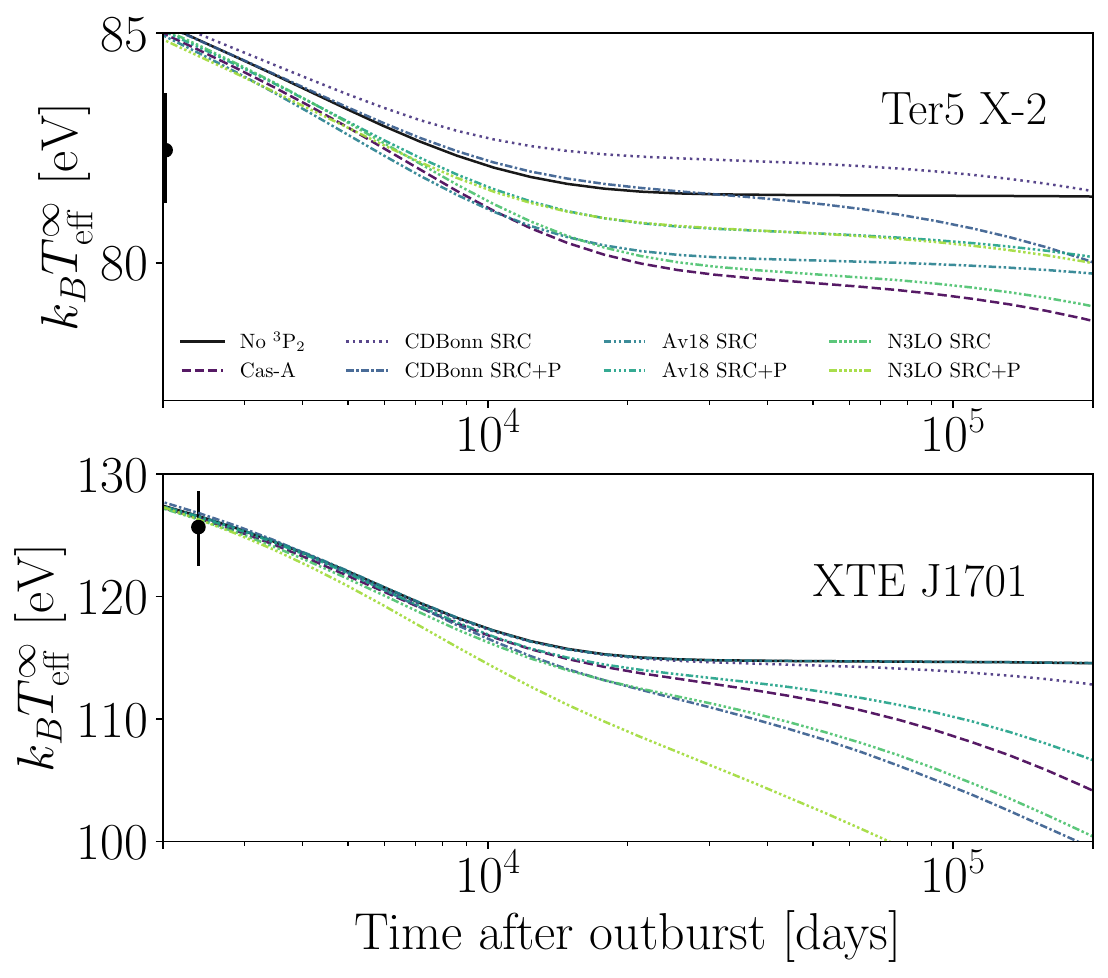}
\caption{Same as Fig.~\ref{fig:best_fitting_curves_1L}, now for \igrbrief{} and \xtebrief.}
\label{fig:best_fitting_curves_2L}
\end{figure}

The results discussed in Sections~\ref{subsec:discuss_1} and~\ref{subsec:discuss_2} motivate a further question: does neutron \superfluidp{} pairing have observable implications for the long-term thermal evolution of these sources? 

To explore this question, we extended the eight cooling simulations for each source to approximately $10^{6}$ days after the end of the outburst. The resulting curves are shown in Figs.~\ref{fig:best_fitting_curves_1L} and~\ref{fig:best_fitting_curves_2L}.
The long-term behavior predicted by these models should be taken as overall trends given that subsequent accretion episodes may alter the actual thermal evolution. This is the case of \mxbbrief{}, a source which in 2015 underwent a second outburst episode \citep[see][]{Parikh2019}, as well as for \xtebrief{} and its 2022 outburst \citep{Iwakiri_2022, Agrawal_2025}. Still, previous works have shown that as long as such outbursts are short-lived, the overall cooling trend established by an earlier outburst is not completely erased \citep{Turlione_2015A&A...577A...5T}. For completeness, we include in the plot for \mxbbrief{} the available cooling data obtained after its second outburst.

\begin{figure}
    \includegraphics[width=1.0\linewidth]{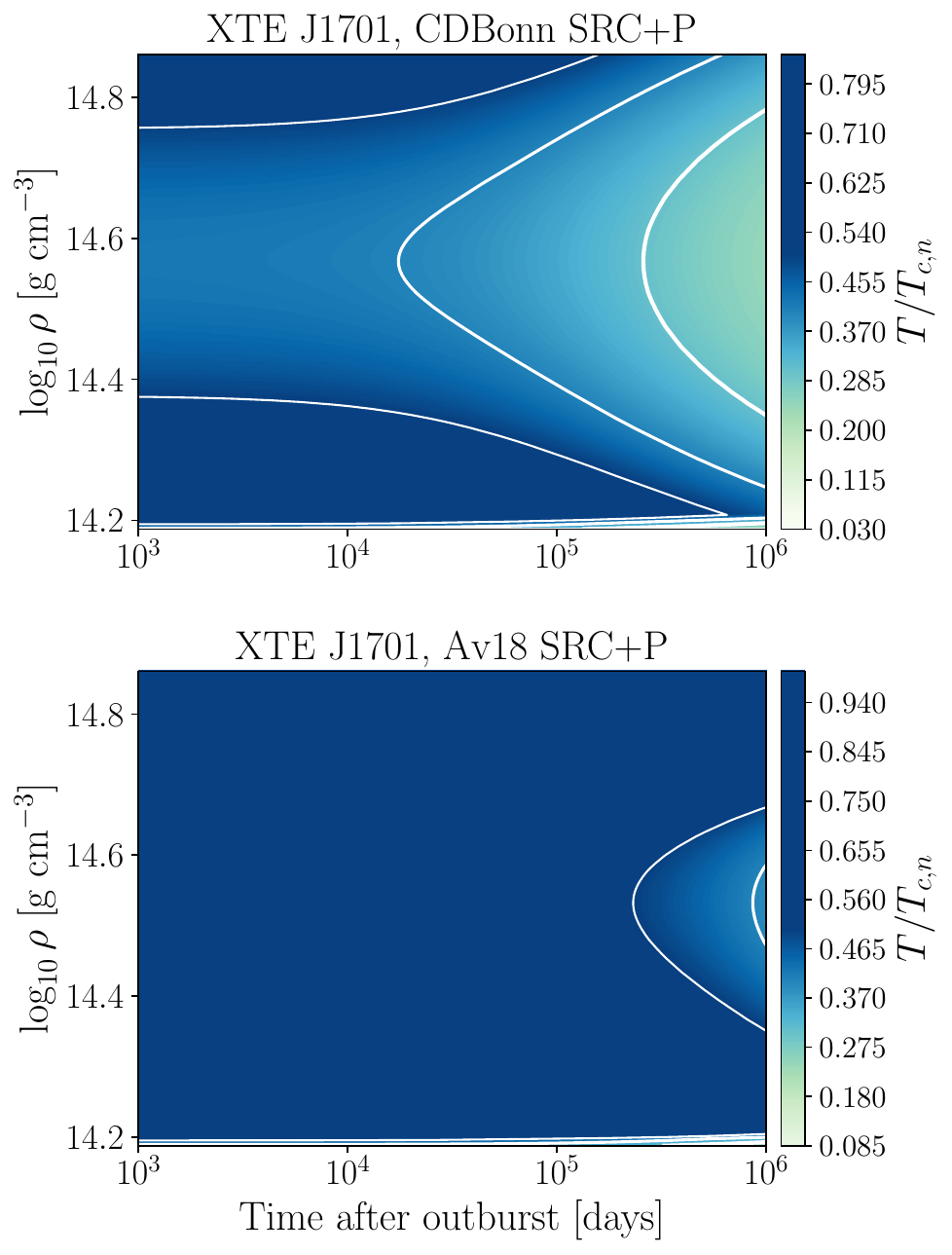}
\caption{Temperature profile $T/T_{c,n}$ as a function of the average density $\rho$ in g~cm$^{-3}$ in the core of the neutron star for the source \xtebrief, considering the \dingii{} and \dingiv{} neutron \superfluidp{} pairing gaps. Time is measured in days since the end of the outburst episode.}
\label{fig:t_tcn_plot_xte}
\end{figure}

On the one hand, for the cool sources \mxbbrief{} and \ksbrief{} differences between the baseline and neutron \superfluidp{} superfluid models become significant only beyond about $3\times 10^{5}$ days after the end of the outburst. The inferred core temperature of \ksbrief{} is roughly twice that of \mxbbrief{}, and its predicted cooling is correspondingly faster in the models considered here. On the other hand, for the hot sources \igrbrief{} and \xtebrief{} the differences between models with and without neutron \superfluidp{} superfluid become apparent at earlier times, around $10^{4}$ days after the end of the outburst. In the case of \xtebrief{}, the three gaps from \citet{Ding_2016} that remain nonzero up to the highest Fermi wavenumbers considered here, namely \dingii{}, \dingv{} and \dingvi{}, predict the fastest cooling among the adopted pairing gaps. In addition, the predicted cooling of \xtebrief{} at $t\geq 10^{4}$ days after its 2013 outburst via PBF neutrino emission is different between the \dingii{} and \dingiv{} gaps due to their closure at high densities, despite both sharing a comparable $T_{c,n,\mathrm{max}}$. To illustrate this point, in Fig.~\ref{fig:t_tcn_plot_xte} we display the evolution of the core's $T/T_{c,n}$ ratio for these two gaps. This behavior suggests that the assumed density dependence of the gap may influence the long-term evolution predicted by the models. Whether these differences can be distinguished observationally will depend on future measurements and on the occurrence of any intervening accretion episodes.

\begin{figure}
    \includegraphics[width=1.0\linewidth]{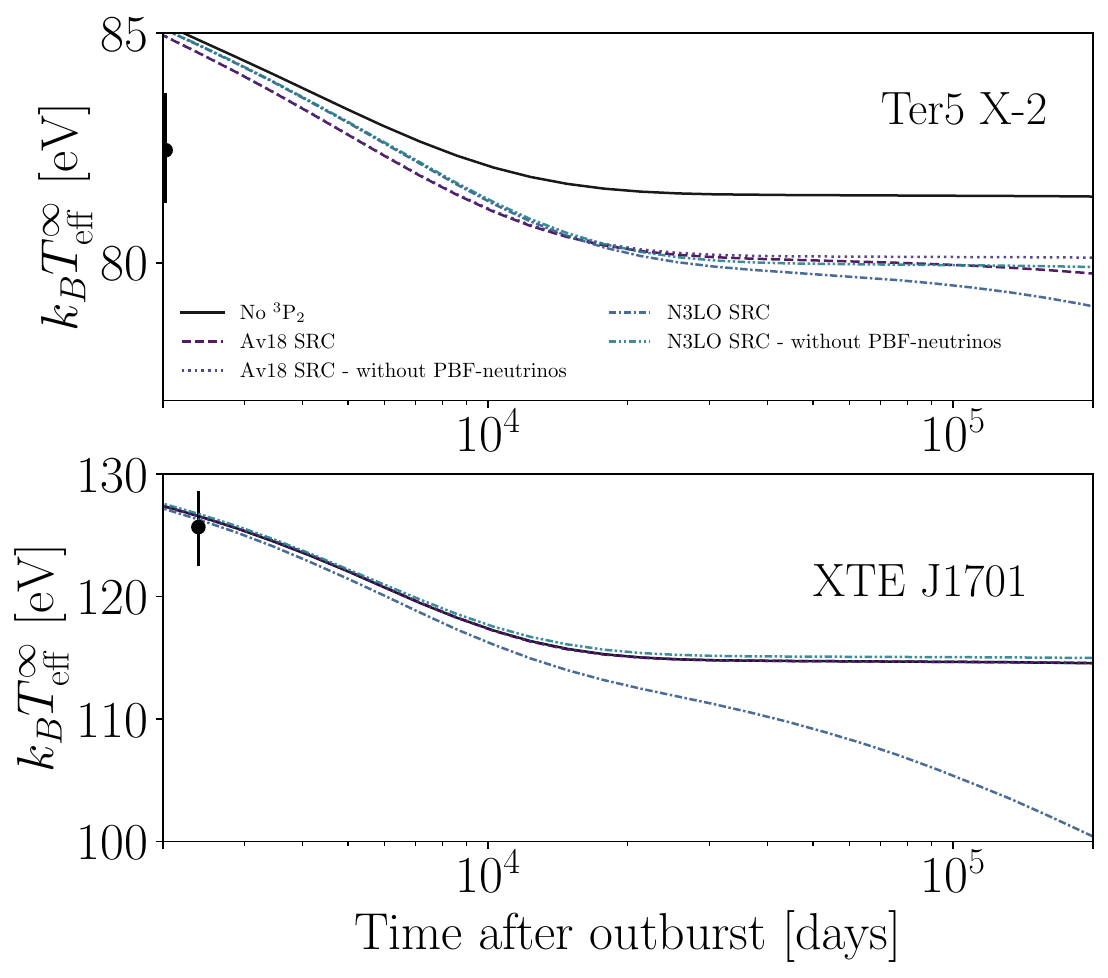}
\caption{Same as Fig.~\ref{fig:best_fitting_curves_1L}, now for \igrbrief{} and \xtebrief, considering the baseline model and cooling curves with and without PBF neutrino emission from the neutron \superfluidp{} channel.}
\label{fig:best_fitting_curves_3L}
\end{figure}

In Section~\ref{subsec:discuss_1}, we observed that the cooling of \igrbrief{} and \xtebrief{}, being the hottest sources, might be affected by PBF neutrino emission rather than by the suppression of the neutron specific heat. To verify it, we carried out additional simulations employing the best-fitting parameters from the MCMC runs but removing the PBF neutrino emission due to the \superfluidp{} channel. In Fig.~\ref{fig:best_fitting_curves_3L}, we display the cooling curves from these simulations for \igrbrief{} and \xtebrief. Let us recall that for \igrbrief{} the best-fitting curve was obtained from the \dingiii{} gap whereas for \xtebrief{} it is from the \dingiv{} gap. In the case of \igrbrief{}, we observe that removing the \superfluidp-PBF neutrino emission results in slightly larger effective temperatures.
In contrast, for \xtebrief{} the models using the \dingiii{} gap with and without \superfluidp{}-PBF neutrino emission lead to almost identical cooling curves; a discrepancy is only seen when using the \dingv{} gap. 
From these simulations, it follows that PBF neutrino emission from the \superfluidp{} channel plays a role in enhancing the cooling at $t\geq 10^{4}$ days after the end of an outburst for these sources. However, the magnitude of such cooling is sensitive to the maximum size of the gap, since $\Delta_\mathrm{max}\sim 10^{-1}$ MeV for \dingv{} whereas it is $\sim 4\times 10^{-2}$ MeV for \dingiii{}.


\subsection{How do we reconcile our results with the rapid cooling of Cassiopeia-A?}\label{subsec:discuss_4}

The results discussed in Sections~\ref{subsec:discuss_1} to~\ref{subsec:discuss_3} suggest that neutron \superfluidp{} pairing can only modestly modify the inferred thermal evolution of transiently accreting neutron stars after an accretion episode, at least up to about $10^4$ days in quiescence. This contrasts with the interpretation of the isolated neutron star Cassiopeia A, whose observed rapid cooling $\sim 300$ years after its birth has been argued as evidence in favor of neutron \superfluidp{} pairing in the core \citep{Shternin_2011MNRAS.412L.108S, Page_2011PhRvL.106h1101P}.

This raises the question of how the two pictures be reconciled.
If the present thermal emission Cassiopeia A were interpreted, purely for comparison, as that of a neutron star observed approximately $10^{5}$ days after a hypothetical early accretion episode, its evolution could be contrasted with here (e.g. Table~\ref{tab:results_all_sources}), potentially with a temperature of order $10^{9}$ K. At such temperatures, the ratio $T/T_{c,n}$ may be of order unity for some regions of the core, depending on the adopted gap. For instance, if $T_{c,n,\mathrm{max}}\sim 0.6$ GK and since the temperature of the star is bound to decay, from Fig.~\ref{fig:neutrino_factors} we see that for $T/T_{c,n}\geq 10^{-1}$ the neutron specific heat is less susceptible to be put into test whereas the neutrino emission plays a more substantial role, first from the \superfluids{} channel and only as the star further cools down from the \superfluidp{} channel. Since the onset of such processes is controlled by $T_{c,n,\mathrm{max}}$, we see why Cassiopeia A is a probe of this parameter. In contradistinction to this scenario, in the cooling of transiently accreting neutron stars we have identified that typical $T/T_{c,n,\mathrm{max}}$ ratios range from $\sim 0.075$ to $\sim 0.625$, depending on the exact value of \tcore{} (see Sect.~\ref{subsec:discuss_1}).

In this limited sense, the rapid cooling predicted for the hot sources \igrbrief{} and \xtebrief{}, between approximately $10^{4}$ and $10^{5}$ days after the end of an outburst (see Fig.~\ref{fig:best_fitting_curves_2L}), offers a qualitative point of comparison with the rapid cooling of Cassiopeia A. Therefore, if upcoming observations show a trend of rapid cooling then we might consider that the presence of neutron \superfluidp{} superfluidity, similarly as in Cassiopeia A, provides a plausible explanation.


\subsection{Is neutron \superfluidp{} pairing essential for cooling of transient sources?}\label{subsec:discuss_5}

For each of the four sources considered in this work, at least one model including neutron \superfluidp{} superfluidity yields a lower $\chi^2$ than the corresponding baseline model without. However, the magnitude of this improvement is small: the difference in $\chi^2$ is at most approximately $0.10$ for \mxbbrief{} and \ksbrief{}, and approximately $0.20$ for \igrbrief{} and \xtebrief{}. These differences are also small when expressed in terms of the reduced $\chi^2$.

Taken together, the results indicate that neutron \superfluidp{} superfluidity provides, at most, a modest improvement to the fits of the four transient sources considered here. The available data do not establish that this pairing channel is essential for reproducing their observed post-outburst cooling curves. At the same time, the results do not imply that it is physically irrelevant: the impact on the fitted curves depends on the source and the assumed gap, and the inferred crustal parameters can vary accordingly. The extended simulations in Figs.~\ref{fig:best_fitting_curves_1L} and~\ref{fig:best_fitting_curves_2L} further show that \superfluidp{} pairing may have a more noticeable effect on the predicted long-term evolution of the hottest sources, particularly \igrbrief{} and \xtebrief{}. This could be tested by future observations.

\section{Conclusions}\label{sect:conclusions}

In this work, we have investigated the post-outburst cooling of neutron stars in four transient sources: \mxb, \ks, \igr{} and \xte, taking into account the effects of neutron \superfluidp{} superfluidity. We considered six pairing gaps based on microscopic calculations, together with the phenomenological gap adopted in previous studies of the rapid cooling of Cassiopeia A. Overall, including neutron \superfluidp{} superfluidity yields only a modest improvement in the quality of the fits relative to the baseline models without. However, data are insufficient for determining the gap despite the diversity of the sources in terms of outburst duration and effective temperature at $\sim 10^4$ days after the end of the outburst. 

Our results also suggest a possible association between the inferred core temperature \tcore{} and the impact of \superfluidp{}  pairing on the fit quality, although the limited number of sources prevents us from establishing a general trend. The extended cooling simulations indicate that the differences between models with and without \superfluidp{} pairing may become more apparent at late times, particularly for the hottest sources, \igrbrief{} and \xtebrief{}.
Future observations of these sources could the different scenarios and better constrain the neutron \superfluidp{} gap. 

The phenomenological Cas-A gap provides an additional point of comparison. Relative to the baseline models, it improves the fit only for \ksbrief{} and \xtebrief{}. Although this gap has not been derived from microscopic calculations, its inclusion allows us to assess the sensitivity of our results to a prescription previously used in neutron-star cooling studies. Its performance should not be interpreted as a strict worst-case scenario for all possible pairing gaps. 

Taken together, our results indicate that the impact of neutron \superfluidp{} superfluidity depends on both the source and the adopted gap. While the present data do not establish  \superfluidp{} pairing as an essential ingredient for reproducing the observed post-outburst cooling curves, the best fits independently support the constraint $T_{c,n,\mathrm{max}} \sim 0.5$ GK obtained from the cooling of Cassiopeia A or the $r$-mode instability of neutron stars in LMXB \citep{Kantor_2020, Shternin_2021}. On the other hand, these astrophysical systems offer an insight regarding the high-density dependence of the gap as our fits tend to favor gaps closing at $k_{F,n}\sim 2.5$ fm$^{-1}$. This motivates further investigation of whether specific gap properties can leave identifiable signatures in the long-term thermal evolution of transiently accreting neutron stars.

Thus, our present calculations yield two principal conclusions, each addressed to a different community. On the nuclear-physics side, our results indicate that the late-time cooling of transiently accreting neutron stars during quiescence may probe the maximum density at which the neutron \superfluidp{} pairing gap closes. This is complementary to, and potentially independent of, the constraint on $\Delta_\mathrm{max}$ inferred from the rapid cooling of the young isolated neutron star in Cassiopeia A. On the observational side, the need for data at $t\geq 10^{4}$ days after the end of an outburst highlights the importance of monitoring campaigns during the quiescent phase of the hottest transient X-ray sources.

\begin{acknowledgements}
This work was supported by the Fonds de la Recherche Scientifique-FNRS under Grant No IISN 4.4502.19.
\end{acknowledgements}

\bibliographystyle{aa}
\bibliography{nscool_3p2}

\begin{appendix}

\section{Comparison of neutron \superfluidp{} pairing gaps}\label{appendix:gaps_comparison}

\begin{figure}
    \centering
    \includegraphics[width=1.0\linewidth]{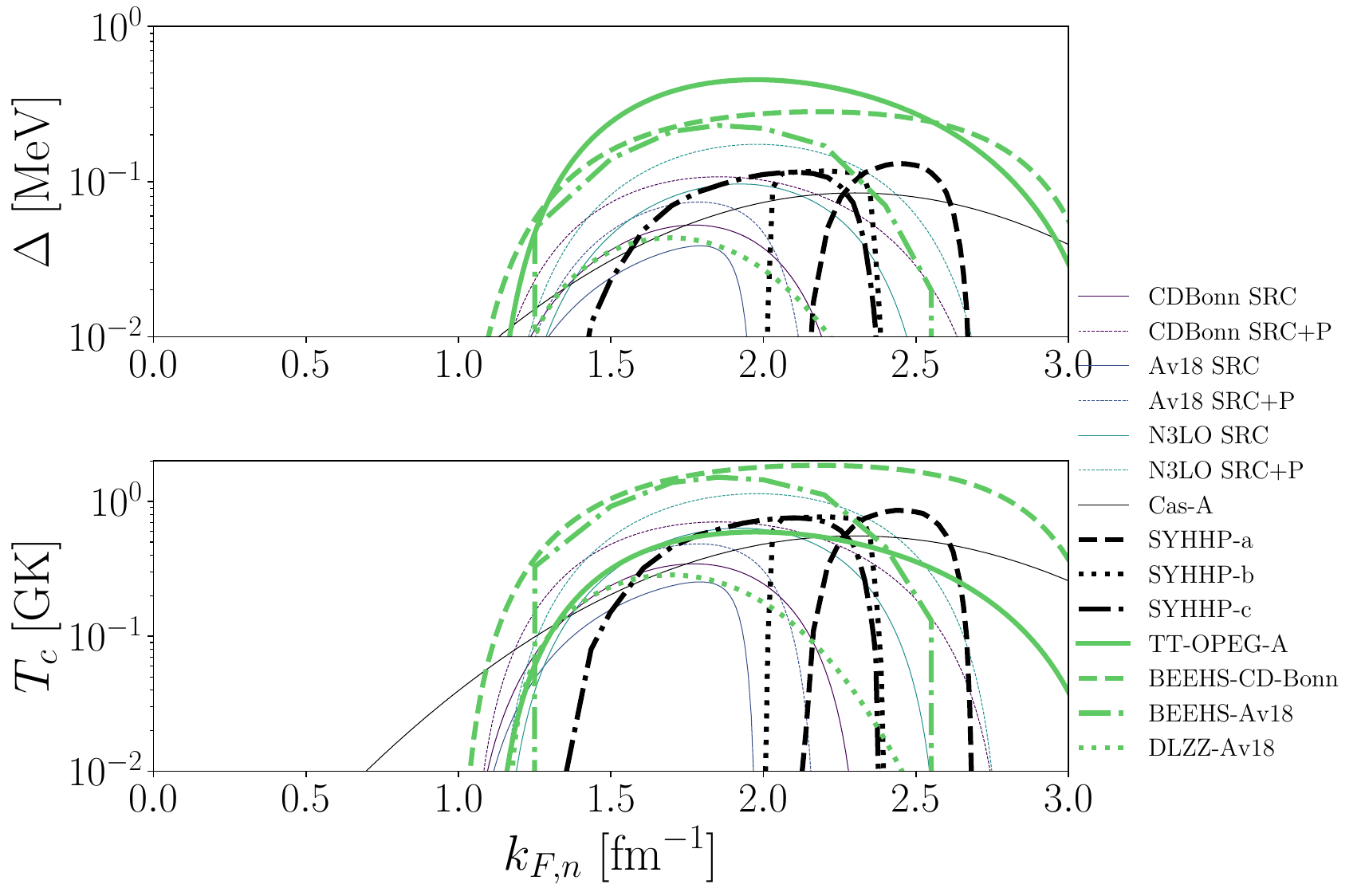}
\caption{Upper panel: Neutron \superfluidp{} pairing gaps employed in the present work as a function of the neutron Fermi wavenumber. Lower panel: corresponding critical temperature $T_{c}$ (in units of GK) for each of the gaps from the upper panel.}
\label{fig:pairing_gaps_2}
\end{figure}

The relation between the pairing gap $\Delta$ and the critical temperature $T_{c}$ for the \superfluids{} pairing gap is given by $k_{B}T\approx 0.5669\Delta$ \citep{Baldo_1992, 1999PhyU...42..737Y, Page_2004}. For the \superfluidp{} channel, however, such relation between $\Delta$ and $T_{c}$ strictly does not hold since, by definition, the generalized \superfluidp{} pairing gap is anisotropic.
In practice, however, the aforementioned relation holds for the neutron \superfluidp{} pairing within the ``angle-averaged approximation'' \citep{Baldo_1992, Takatsuka_1993, Page_2004, Page_2011PhRvL.106h1101P, Ding_2016}. This consists in replacing the bare $\Delta$ in the above expression by 
$\sqrt{\langle D^{2}\rangle}$, i.e. the square root of the angle-averaged term $D^{2}$ appearing in the quasiparticle energy.
\citet{Baldo_1992} stated the numerical validity of such approximation and stressed on reporting $\sqrt{\langle D^{2}\rangle}$ in the summary of their results. 
With respect to this angle-averaged approximation, \citet{Takatsuka_1993} points out that the ``formal'' relation should read as $k_{B}T_{c}$ $\approx 0.57\frac{\sqrt{8\pi}}{\sqrt{2}\Gamma_{0}}$ $\sqrt{\langle D^{2}\rangle}$, where $\ln\Gamma_{0} = 1.22$. Upon simplification, $k_{B}T_{c}\approx 0.59\sqrt{\langle D^{2}\rangle}$, in rather good agreement with \citet{Baldo_1992}.
Regarding the pairing gaps considered in this work, \citet{Ding_2016} states to be using this angle-averaged approximation for their reported gaps. This can also be corroborated by their statement about the compatibility of their results with the equivalent $\Delta$ of the $T_{c}$ reported in \citet{Page_2011PhRvL.106h1101P}, e.g. $\sim 0.08$ MeV (see Fig.~\ref{fig:pairing_gaps}). Consequently, for converting all their gaps into $T_{c}$, it suffices to use $k_{B}T\approx 0.5669\Delta^\mathrm{Ding}$, where $\Delta^\mathrm{Ding}$ is the fitting formulae reported in this paper.

Having clarified the convention in literature, let us now compare the gap sizes and critical temperatures employed in this work with others from literature. In Fig.~\ref{fig:pairing_gaps_2} we display such comparison.
\begin{itemize}
    \item The gaps which we label as SYHHP-a, SYHHP-b and SYHHP-c correspond to the phenomenological models for $T_{c}(k_{F,n})$ by \citet{Shternin_2011MNRAS.412L.108S}. 
    These authors report $T_{c}(\rho)$, having converted $k_{F,n}$ into $\rho$ via one version of the APR-EOS from \citet{Akmal:1998vu}, which produces neutron star maximum mass of $1.929 M_{\odot}$. Such APR-EOS, however, is different from the APR-EOS considered by \citet{Page_2011PhRvL.106h1101P} as in this latter work the maximum mass reached is of $2.18 M_{\odot}$. In the absence of a published expression of $T_{c}(k_{F})$, our conversion rules are thus an educated guess. To convert these phenomenological $T_{c}$ into $\Delta$, we consider the first conversion rule, in contrast to \citet{Ho_2015} who considered the second conversion rule. As a consequence, all a,b and c models from \citet{Shternin_2011MNRAS.412L.108S} result in pairing gaps of the size of $\sim 10^{-1}$ MeV.
    \item The TT-OPEG-A gap comes from the calculations by \citet{Takatsuka_2004} employing the OPEG-A potential. The fitting formula was taken from \citet{Ho_2015}. These authors considered a large collections of gaps for their study of the neutron star cooling of Cassiopeia A. Among these, the TT-OPEG-A - there labelled as TToa - adopts the most relevant role as it is the one used in their best-fit model. This is also the gap employed in the hybrid cooling scenario by \citet{Leinson_2022}.
    \item \citet{Baldo_1998} calculated the neutron \superfluidp{} pairing for a collection of nucleon potentials.
    Among these, \citet{Ho_2015, Shternin_2021} label as BEEHS the gap corresponding to the CD-Bonn potential, whereas \citet{Sedrakian_2019} label as GM the gap corresponding to the Av18 potential. For the present comparison and visual clarity we opt for labeling these gaps as BEEHS-CD-Bonn and BEEHS-Av18 respectively. For the first, we use the fitting formula from \citet{Ho_2015} whereas for the second one we extract the data directly from \citet{Baldo_1998}. These gaps have $\Delta_\mathrm{max} \approx 0.3$ MeV, and consequently $T_{c,n,\mathrm{max}}\sim 1$ GK.
    \item The gap labelled as DLZZ-Av18 was calculated by \citet{Dong_2016} using the Av18 potential.
    For the plot, we use their fitting formula reported in \citet{Dong_2021}. Since they fit the angle-averaged gap as a function of the baryon number density in fm$^{-3}$, for illustrative purposes we use their fitting formula assuming pure neutron matter, e.g. $k_{F} = (3\pi^{2}n_{b})^{1/3}$. Incidentally, we note that \citet{Sedrakian_2019} label as GMZ the calculations by \citet{Dong_2013} in which they use the same formalism as in \citet{Dong_2016} but with the Bonn-B potential. However, the difference between these gaps is relatively small.
\end{itemize}
With respect to the Cas-A and SYHHP-a, SYHHP-b and SYHHP-c gaps, we observe that these phenomenological gaps agree well with respect to $T_{c,\mathrm{max}}$, although Cas-A covers a larger region in $k_{F,n}$ than these three other gaps. With respect to other microphysical calculations, we observe that all gaps by \citet{Ding_2016} predict systematically smaller $\Delta_\mathrm{max}$ than the old TT-OPEG-A and BEEHS-CD-Bonn and BEEHS-Av18 gaps. In contrast, the DLZZ-Av18 gap predicts a rather small value for $\Delta_\mathrm{max}$, albeit its value is comparable to the calculations by \citet{Ding_2016} using the Av18 potential.


\section{The effects of neutron \superfluidp{} pairing over the neutron specific heat and the PBF emission}\label{appendix:3p2_r_f}

Within the BCS formalism, neutron superfluidity - either in the \superfluids{} or in the \superfluidp{} channel - changes the neutron specific heat via a suppression factor $R$, i.e. $c_{P, n} = R(T/T_{c,n})\, c^{\text{normal}}_{P, n}(T)$, where $c^{\text{normal}}_{P, n}(T)$ represents the standard specific heat for neutrons(fermions) \citep{1999PhyU...42..737Y}. The $R$ factor for both channels is shown in Fig.~\ref{fig:neutrino_factors}.
The neutrino emissivity due to PBF, either in the \superfluids{} or in the \superfluidp{} channel, can be expressed as \citep{1999PhyU...42..737Y}
\begin{equation}
Q = 1.170 \times 10^{21}\left(\frac{m^{\ast}_{N}}{m_{N}}\right)\left(\frac{p_{F}}{m_{N}c}\right)
T_{9}^{7}\mathcal{N}_{\nu}\, a\, F
\end{equation}
where $T_{9}=T/(10^{9}\mathrm{K})$, $a$ is a numerical factor, $\mathcal{N}_\nu$ denotes the neutrino flavors, $m^{\ast}_{N}/m_{N}$ is the ratio of effective to rest nucleon mass, $p_{F}$ is the neutron Fermi momentum and $F$ represents the dimensionless function arising from the phase-space integrals and depends on the ratio $T/T_{c}$. This $F$-function is shown in the lower panel of Fig.~\ref{fig:neutrino_factors} for both \superfluids{} and \superfluidp{} channels.


\end{appendix}

\end{document}